\documentstyle[prd,aps,preprint,tighten,floats]{revtex}
\begin{document}
\preprint{CERN-TH/99-101, hep-th/9904086}
\date{April, 1999}
\title{Generalized Conformal Quantum Mechanics of D0-brane}
\author{Donam Youm
\thanks{E-mail address: Donam.Youm@cern.ch}}
\address{Theory Division, CERN, CH-1211, Geneva 23, Switzerland}
\maketitle
\begin{abstract}
We study the generalized conformal quantum mechanics of the probe D0-brane 
in the near horizon background of the bound state of source D0-branes.  
We elaborate on the relationship of such model to the M theory in the 
light cone frame.
\end{abstract} 

\section{Introduction}

According to the holographic principle \cite{hoof,suss1,suss2}, the bulk 
theory (with gravity) and the boundary theory (without gravity) are
equivalent.  Such correspondence gives insight into one theory
from the other.  Interest in the holographic principle is 
revived from the recent observation \cite{mald1} that the bulk 
gravity theory in the near horizon geometry of brane configuration 
is equivalent to the boundary theory described by the corresponding 
worldvolume field theory in the decoupling limit.  (The closely related 
previous works in Refs. \cite{duff1,town1,town2} study the correspondence 
between the bulk theory on the AdS space and the boundary supersingleton field 
theory.)  In particular, when the near horizon geometry is the AdS space, 
i.e. D3-brane, M2-brane and M5-brane cases, the boundary theory is conformal.
However, it is found out in Refs. \cite{jevi1,jevi2} that even for other 
branes, whose near horizon geometry is not the AdS type and therefore 
the corresponding boundary theory is not a ``genuine'' conformal
theory, one can still define ``generalized'' conformal theory.  

The symmetry group of such generalized conformal theory can be manifestly 
seen in the bulk theory in the so-called ``dual'' frame 
\cite{duff2,town1,town2,boon1,boon2,town3}.  The dual frame is regarded as a 
preferred frame for supergravity probes in the near horizon background of 
the source branes.  In this frame, the near horizon geometry of a $p$-brane 
supergravity solution takes the AdS$_{p+2}\times S^n$ form.  The $SO(p+1,2)$ 
isometry of the AdS$_{p+2}$ part is realized in the boundary theory as the 
``generalized'' conformal symmetry of Refs. \cite{jevi1,jevi2}.  The dual 
frame is also called a ``holographic'' frame, since an UV/IR connection 
between the bulk and the boundary theory is manifest in this frame. 

In this paper, we study the D0-brane case of such generalized conformal 
theory.  D0-brane is particularly interesting for its relevance to M 
theory.  An important discovery of the M(atrix) model
\footnote{This model was proposed \cite{halp1} long ago as the $N=16$ 
supersymmetric gauge quantum mechanics.  See also Ref. \cite{halp2}.}
is that difficult problems of quantum M theory is reduced to non-relativistic 
quantum mechanics.  Originally, it is conjectured \cite{matr1} that M-theory 
in the infinite momentum frame (IMF) \cite{wein} is exactly described by the 
$U(\infty)$ $D=1$ super-Yang-Mills (SYM) theory, which is the worldvolume 
theory of the bound state of the infinite number ($N\to\infty$) of D0-branes.  
The $D=1$ $U(N)$ SYM theory, which is the supersymmetric matrix model 
description of the supermembrane theory \cite{dewi}, is nothing but the 
maximally supersymmetric $U(N)$ Yang-Mills theory dimensionally reduced 
from $9+1$ to $0+1$ dimensions (i.e. $N\times N$ hermitian matrix quantum 
mechanics).  It is further conjectured \cite{matr2} that the equivalence 
of (M)atrix theory to M-theory is also valid for a finite $N$.  
The conjecture states that M-theory compactified on a light-like circle 
with finite momentum along the circle is exactly described by a $U(N)$ 
matrix theory.  The quantization of such theory is called the discrete 
light-cone quantization (DLCQ).  One can think of this light-like circle 
as a small space-like circle that is boosted by a large amount \cite{seib1}.  
Just like M-theory in the IMF, M-theory in the light-cone frame (LCF) 
is described purely by D0-branes with positive momentum and has the 
Galilean invariance in the transverse space \cite{matr2}.  Therefore, 
M-theory on a light-like circle can be viewed as a theory of the finite 
number of D0-branes in the low velocity (non-relativistic) limit.  

This idea can also be understood from the bulk/boundary duality as 
follows.  When compactified on a light-like circle, the supergravity 
M-wave solution (viewed as the supergraviton with the momentum number 
$N$) becomes the near horizon limit of supergravity solution for $N$ 
coinciding D0-branes \cite{tsey1}.  Since M-theory in the LCF in the 
supergravity solution level is related to the near horizon limit of 
the supergravity D0-brane solution, by the generalized AdS/CFT 
duality \cite{mald2,jevi1,jevi2} M-theory in the LCF has to be related to 
the boundary theory of the bound state of $N$ D0-branes. This boundary theory 
is the maximally supersymmetric $SU(N)$ Yang-Mills theory dimensionally 
reduced from $9+1$ to $0+1$ dimensions, as stated in the (M)atrix theory 
conjecture.

In this paper, we view the non-relativistic (matrix) quantum mechanics 
of bound states of D0-brane as a system of the probe D0-brane moving 
in the background of large number of source D0-brane bound state.  
This view of (M)atrix model is also taken in Ref. \cite{tsey1}, which 
reproduces the (M)atrix model graviton-graviton scattering calculation 
from the effective action for a probe moving in the background of the 
M-wave supergravity solution.  It is further shown in Refs. \cite{jevi1,jevi2} 
that the Dirac-Born-Infeld (DBI) action for a radially moving probe 
D0-brane in the background of the source D0-brane bound state can also be 
determined by imposing the generalized conformal symmetry of the boundary 
theory, which is just the $D=1$ $U(N)$ SYM theory describing the 
M theory in the LCF.

The quantum mechanics of the probe D0-brane in the near horizon 
background of the source D0-brane is a reminiscence of the conformal 
quantum mechanics of the charged test particle moving in the near 
horizon background of the $D=4$ extreme Reissner-Nordstr\"om black 
hole studied in Ref. \cite{kall1}.  The radial motion of such test particle 
is described by the relativistic conformal quantum mechanics with the 
$SL(2,{\bf R})\cong SU(1,1)$ symmetry.  This is a generalized version 
of (non-relativistic) conformal mechanics studied in Refs. 
\cite{scm1,ap,scm2}.  One can view such generalized conformal mechanics 
of the radial motion of the test particle as the boundary conformal field 
theory counterpart of the bulk gravity theory in the near horizon Ads$_2$ 
geometry of the extreme Reissner-Nordstr\"om black hole, since the $SO(1,2)
\cong SU(1,1)$ isometry of the AdS$_2$ space is realized as the symmetry 
of the dynamics of the test particle in the spacetime with one lower 
dimension.  Since the near horizon geometry of the source D0-brane 
supergravity solution in the dual frame is AdS$_2\times S^8$, one 
would expect that the dynamics of the probe D0-brane in this 
background has the $SL(2,{\bf R})\cong SO(2,1)$ symmetry.  In the D0-brane 
case, since this symmetry does not extend to the genuine conformal symmetry 
of the boundary theory but only to the so-called generalized conformal 
symmetry \cite{jevi1,jevi2} as pointed out in the above, the quantum 
mechanics of the probe D0-brane will have a generalized conformal symmetry.  

The paper is organized as follows.  In section 2, we summarize the 
properties of the near horizon geometry of the supergravity D0-brane
solution.  In section 3, we study the ``generalized'' conformal 
quantum mechanics of D0-brane, elaborating its relation to M theory 
as pointed out in the above. 

\section{$D0$-brane Solution in the Near Horizon Region}

In this section, we survey aspects of the $D0$-brane supergravity 
solution and its near horizon geometry, illuminating their relations. 

The string-frame type-IIA effective supergravity action for $D0$-brane 
solution is given by
\begin{equation}
S={1\over{16\pi G_{10}}}\int d^{10}x\sqrt{-G^{str}}
[e^{-2\phi}({\cal R}^{str}+2\partial_M\phi\partial^M\phi)
-{1\over{2\cdot 2!}}F_{MN}F^{MN}],
\label{iibsg}
\end{equation}
where $G_{10}$ is the 10-dimensional gravitational constant, 
${\cal R}^{str}$ and $G^{str}$ are the Ricci scalar and the 
determinant for the the string-frame metric tensor $G^{str}_{MN}$, 
$\phi$ is the dilaton in the NS-NS sector and $F_{MN}$ is the field 
strength for the RR 1-form potential $A_M$ that $D0$-branes couple to.  

The supergravity solution for the $D0$-brane has the following form: 
\begin{eqnarray}
ds^2_{str}&=&G^{str}_{MN}dx^Mdx^N=-H^{-{1\over 2}}dt^2+H^{1\over 2}
(dx^2_1+\dots+dx^2_9),
\cr
e^{\phi}&=&g_{s}H^{3\over 4},\ \ \  A_t=-g^{-1}_{s}H^{-1};\ \ \ 
H=1+{Q\over{r^7}}\equiv 1+\left({{\mu}\over r}\right)^7,
\label{sgdzero}
\end{eqnarray}
where $g_s$ is the string coupling constant, which is just the vacuum 
expectation value or the asymptotic value of the dilaton $e^{\phi}$.
Here, the $D0$-brane charge $Q$ is related to the string theory 
quantities as
\begin{equation}
Q=(\alpha^{\prime})^{7\over 2}g_sN,
\label{charge}
\end{equation}
where $\alpha^{\prime}$ is related to the string length scale $l_s$ 
as $l_s=\sqrt{\alpha^{\prime}}$ and $N$ is the number of the $D0$-branes. 
By expanding the Dirac-Born-Infeld (DBI) action for the D0-brane to the 
lowest order in $\alpha^{\prime}$, one finds the following relation of 
the string theory quantities to the Yang-Mills gauge coupling $g_{YM}$:
\begin{equation}
g^2_{YM}={{g_s}\over{(\alpha^{\prime})^{3\over 2}}}.
\label{ymcouple}
\end{equation}
Therefore, the constant $Q$ in the harmonic function is rewritten in 
terms of the SYM quantities as $Q=\mu^{7}=(\alpha^{\prime})^5g^2_{YM}N$. 

In order to decouple the massive string modes (whose masses are 
proportional to $1/\alpha^{\prime}$) and the gravity modes (whose 
strength goes as $g^2_s(\alpha^{\prime})^4$) from the massless 
open string modes, which describe the Yang-Mills theory, one has to 
take $\alpha^{\prime}\to 0$, while keeping $g_{YM}$ as a finite constant, 
which means for the D0-brane case $g_s\to 0$ as well (Cf. Eq. 
(\ref{ymcouple})).  Furthermore, the above metric (\ref{sgdzero}) can be 
regarded as the gravitational field felt by a probe D0-brane in the 
background of the collection of $N$ numbers of source D0-branes with the 
radial coordinate $r$ being interpreted as the distance between and probe 
and source $D0$-branes \cite{mald3}.  So, in order to keep the mass 
$m_{string}=r/\alpha^{\prime}$ of the state of open string, which stretches 
between the source and the probe, finite, one has also take the limit 
$r\to 0$ while keeping the following combination as a finite constant:
\begin{equation}
U\equiv {r\over{\alpha^{\prime}}},
\label{higgs}
\end{equation}
thereby going to the near horizon region of the supergravity solution 
(\ref{sgdzero}).  This combination also corresponds to the conventional 
super-Yang-Mills scalar $\Phi^I=X^I/l^2_s$, whose vacuum expectation 
value sets the energy scale.  In this decoupling limit, the supergravity 
solution (\ref{sgdzero}) takes the following form:
\begin{eqnarray}
ds^2_{str}&=&-\left({r\over\mu}\right)^{7\over 2}dt^2+
\left({\mu\over r}\right)^{7\over 2}(dr^2+r^2d\Omega^2_8)
\cr
&=&\alpha^{\prime}\left[-{{U^{7\over 2}}\over{g_{YM}\sqrt{N}}}dt^2
+{{g_{YM}\sqrt{N}}\over{U^{7\over 2}}}(dU^2+U^2d\Omega^2_8)\right],
\cr
e^{\phi}&=&g_{s}\left({\mu\over r}\right)^{{21}\over 4}=g^2_{YM}
\left({{g^2_{YM}N}\over{U^7}}\right)^{3\over 4},
\cr
A_t&=&g^{-1}_{s}\left({r\over\mu}\right)^7={\sqrt{\alpha^{\prime}}
\over{g^2_{YM}}}{{U^7}\over{g^2_{YM}N}}.
\label{sgdecoupl}
\end{eqnarray}

The above supergravity solution (\ref{sgdecoupl}) can be trusted when 
the spacetime curvature ${\cal R}\sim U^3/(g^2_{YM}N)$ is much smaller 
than the string scale $(\alpha^{\prime})^{-1}$ and string coupling 
$e^{\phi}$ is very small, leading to the following constraints on $U$ 
and the original radial coordinate $r$:
\begin{equation}
g^{2\over 3}_{YM}N^{1\over 7}\ll U \ll g^{2\over 3}_{YM}N^{1\over 3},
\label{ucnstrnt}
\end{equation}
\begin{equation}
\sqrt{\alpha^{\prime}}g^{1\over 3}_{s}N^{1\over 7}\ll r \ll 
\sqrt{\alpha^{\prime}}g^{1\over 3}_{s}N^{1\over 3}.
\label{rcnstrnt}
\end{equation}
On the other hand, the near horizon condition $r\ll\mu=Q^{1\over 7}$ 
is expressed in terms of string theory quantities as 
$r\ll \sqrt{\alpha^{\prime}}(g_sN)^{1\over 7}$.  So, for sufficiently 
large $N$ and small $g_s$, the supergravity solution (\ref{sgdecoupl}) 
can be trusted in the overlapping region of $U$.  

Note, the near horizon geometry of the D0-brane supergravity solution 
in the string frame is not AdS$_2\times S^8$, since when the metric is 
expressed in the following suggestive form
\begin{equation}
ds^2_{str}=\alpha^{\prime}\left[-{{U^2}\over\sqrt{\rho_0}}dt^2+
\sqrt{\rho_0}{{dU^2}\over{U^2}}+\sqrt{\rho_0}d\Omega^2_8\right],
\label{adsform}
\end{equation}
the radius $\rho_0\equiv Q/((\alpha^{\prime})^5U^3)$ of the 
would-be AdS space depends on the coordinate $U$.  

However, with the suitable choice of frame, called the ``dual'' frame 
\cite{duff2,town1,town2,town3}, the spacetime metric in the near-horizon 
region takes the AdS$_2\times S^8$ form.  Namely, if one applies the Weyl 
transformation
\footnote{In this paper, we do not include a factor involving $N$ in 
the Weyl transformation of the metric, unlike Ref. \cite{town3}, so that  
the metric in the standard AdS form in the horospherical coordinates has 
explicit dependence on $N$.} 
on the metric as $G^{str}_{MN}\to G^{dual}_{MN}=e^{-{2\over 7}\phi}
G^{str}_{MN}$, then the metric in (\ref{sgdecoupl}) transforms to
\begin{eqnarray}
ds^2_{dual}&=&G^{dual}_{MN}dx^Mdx^N=g^{-{2\over 7}}_{s}\left[
-\left({r\over\mu}\right)^5dt^2+\left({{\mu}\over r}\right)^2dr^2+
\mu^2d\Omega^2_8\right]
\cr
&=&N^{2\over 7}\alpha^{\prime}\left[-{{U^{5}}\over{g^{2}_{YM}N}}dt^{2}
+{{dU^{2}}\over{U^{2}}}+d\Omega^{2}_{8}\right].
\label{dualmet}
\end{eqnarray}
By further redefining the radial coordinate as $\bar{r}=({{25}\over 4}
g^{4\over 7}_{s}\mu^3)^{-{1\over 2}}r^{5\over 2}$ or $u=({{25}\over 4}
g^{2}_{YM}N^{3\over 7})^{-{1\over 2}}U^{5\over 2}$, 
one can bring the metric to the following standard AdS$_2\times S^8$ 
form in the horospherical coordinates:
\begin{eqnarray}
ds^2_{dual}&=&-\left({{5g^{1\over 7}_{s}}\over{2\mu}}\right)^2\bar{r}^2dt^2+
\left({{2\mu}\over{5g^{1\over 7}_{s}}}\right)^2{{d\bar{r}^2}\over{\bar{r}^2}}
+\left({\mu\over {g^{1\over 7}_{s}}}\right)^{2}d\Omega^2_8
\cr
&=&\alpha^{\prime}\left[-\left({{5}\over{2N^{1\over 7}}}\right)^2u^2dt^2+
\left({{2N^{1\over 7}}\over{5}}\right)^2{{du^2}\over{u^2}}+N^{2\over 7}
d\Omega^2_8\right],
\label{dualads}
\end{eqnarray}
where $u=\bar{r}/\alpha^{\prime}$.  
In the dual frame, in which the metric in the near horizon takes the 
above AdS form, the effective action (\ref{iibsg}) takes the 
following form:
\begin{equation}
S={1\over{16\pi G_{10}}}\int d^{10}x\sqrt{-G^{dual}}
\left[e^{-{6\over 7}\phi}({\cal R}^{dual}+{{16}\over {49}}
\partial_M\phi\partial^M\phi)-{1\over 4}e^{{6\over 7}\phi}
F_{MN}F^{MN}\right].
\label{dualact}
\end{equation}

On the other hand, one can view the $D0$-brane solution (\ref{sgdzero}) 
as being magnetically charged under the Hodge-dual field strength to the 
field strength of the 1-form potential in the RR sector.  Namely, the 
supergravity solution (\ref{sgdzero}) solves the equations of motion of 
the following effective action and is magnetically charged under the 
7-form potential $A_{M_1\cdots M_7}$:
\begin{equation}
S^{\prime}={1\over{16\pi G_{10}}}\int d^{10}x\sqrt{-G^{str}}
[e^{-2\phi}({\cal R}^{str}+2\partial_M\phi\partial^M\phi)
-{1\over{2\cdot 8!}}F_{M_1\cdots M_8}F^{M_1\cdots M_8}],
\label{hdgact}
\end{equation}
where $F_{M_1\cdots M_8}$ is the field strength of the 7-form potential 
$A_{M_1\cdots M_7}$.
In the dual frame with the spacetime metric $G^{dual}_{MN}=e^{-{2\over 7}
\phi}G^{str}_{MN}$, the action (\ref{hdgact}) takes the following form 
\cite{town3}:
\begin{equation}
S^{\prime}={1\over{16\pi G_{10}}}\int d^{10}x\sqrt{-G^{dual}}
e^{-{6\over 7}\phi}\left[{\cal R}^{dual}+{{16}\over {49}}
\partial_M\phi\partial^M\phi-{1\over {2\cdot 8!}}
F_{M_1\cdots M_8}F^{M_1\cdots M_8}\right].
\label{dualhdgact}
\end{equation}

As in Ref. \cite{town3}, if one properly includes the factor involving $N$ 
in the Weyl transformation of the metric, i.e. $G^{str}_{MN}\to 
G^{dual}_{MN}=(Ne^{\phi})^{-{2\over 7}}G^{str}_{MN}$, then the 
new radial coordinate $u$ (with which the near horizon metric takes 
the AdS form in the horospherical coordinates) is related to $U$ as 
$u={2\over 5}{{U^{5/2}}\over{g_{YM}N^{1/2}}}$.  This is reminiscence of 
the holographic UV/IR connection between the bulk and the boundary theory 
\cite{suss3,polc1}, if one identifies $u$ with the energy scale $E$ of the 
boundary theory \cite{town3}.  

In this case, the effective action in the dual frame takes the 
following form \cite{town3}:
\begin{equation}
S^{\prime}={{N^2}\over{16\pi G_{10}}}\int d^{10}x\sqrt{-G^{dual}}
(Ne^{\phi})^{-{6\over 7}}\left[{\cal R}^{dual}+{{16}\over {49}}
\partial_M\phi\partial^M\phi-{1\over {2\cdot 8!}}{1\over{N^2}}
F_{M_1\cdots M_8}F^{M_1\cdots M_8}\right].
\label{orgdualact}
\end{equation}
The near horizon form of D0-brane solution to the associated equations 
of motion is:
\begin{eqnarray}
ds^2_{dual}&=&-{{u^2}\over{u^2_0}}dt^2+u^2_0{{du^2}\over{u^2}}
+d\Omega^2_8,
\cr
e^{\phi}&=&{1\over N}(g^2_{YM}N)^{7\over{10}}\left({u\over{u_0}}
\right)^{-{{21}\over{10}}},
\cr
F_8&=&7N\,vol(S^8),
\label{dualneard0}
\end{eqnarray}
where $u_0=2/5$.  From this solution, one can see that there is the 
Freund-Rubin compactification \cite{freu} on $S^8$ of the $D=10$ action 
(\ref{orgdualact}) to the following 2-dimensional effective gauged 
supergravity action \cite{town3}:
\begin{equation}
S^{\prime}=N^2\int d^2x\sqrt{-g}(Ne^{\phi})^{-{6\over 7}}
\left[{\cal R}+{{16}\over{49}}\partial_{\mu}\phi\partial^{\mu}\phi
+{{63}\over 2}\right].
\label{2dact}
\end{equation}
The near-horizon D0-brane supergravity solution (\ref{dualneard0}) is reduced 
under this $S^8$ compactification to a domain solution 
\cite{pope1,pope2,town4}, which is supported by the cosmological term in 
the action (\ref{2dact}).  

\section{Generalized Conformal Mechanics of D0-branes}

We consider a probe D0-brane with mass $m$ and charge $q$ moving in the 
near horizon geometry of the source $N$ D0-brane bound state.  The 
gravitational field felt by the probe D0-brane in the string [dual] frame 
is given by Eqs. (\ref{sgdecoupl}) and (\ref{adsform}) [Eqs. (\ref{dualmet}) 
and (\ref{dualads})].  Here, once again, $r$ is the radial distance between 
the source and the probe D0-branes.  The probe D0-brane moves with 
the 10-momentum $p=(p_M)=(p_t,p_1,...,p_9)$ and its time-component 
is the (static-gauge) Hamiltonian $H=-p_t$.  The expression for 
the Hamiltonian of the probe D0-brane in the near horizon background 
of the source D0-brane can be obtained by solving the mass-shell 
constraint of the probe D0-brane.  

Unlike the case of Ref. \cite{kall1}, which studies a charged particle 
in the Einstein-Maxwell theory, the probe D0-brane satisfies the mass-shell 
constraint which is different from the ordinary constraint $0=(p-qA)^2+m^2=
G^{MN}(p_M-qA_M)(p_N-qA_N)+m^2$.  This is due to the non-trivial dilaton 
field that is present for the D0-brane solution.  So, here we rederive the 
mass-shell condition for the case of D0-branes.  The action for the probe 
D0-brane with the mass $m$ and the charge $q$ moving in the background of 
the source D0-brane has the following form:
\begin{equation}
S=\int d\tau\,L=\int d\tau\left(me^{-\phi}\sqrt{-G^{str}_{MN}\dot{x}^M
\dot{x}^N}-q\dot{x}^MA_M\right),
\label{dpartact}
\end{equation}
where $v^M=\dot{x}^M\equiv{{dx^M}\over{d\tau}}$ is the 10-velocity of the 
probe D0-brane.  Note, since we choose the static gauge for the action, 
the worldline time $\tau$ and the target-space time $t$ are set to equal.  
Once again, $G^{str}_{MN}$ and $A_M$ are the fields produced by the source 
D0-brane.  Note, in this action the metric $G^{str}_{MN}$ is in the string 
frame.  The (generalized) momentum conjugate to $x^M(\tau)$ is
\begin{equation}
P_M=-{{\delta L}\over{\delta\dot{x}^M}}={{me^{-\phi}\dot{x}_M}
\over\sqrt{-G^{str}_{MN}\dot{x}^M\dot{x}^N}}+qA_M.
\label{conjmom}
\end{equation}
As usual, $p_M=m\dot{x}_M/\sqrt{-G^{str}_{MN}\dot{x}^M\dot{x}^N}$ is the 
ordinary 10-momentum of the D0-brane.  From this, one obtains the following 
mass-shell constraint for the probe D0-brane in the string-frame background 
of the probe D0-brane:
\begin{equation}
G^{str\,MN}(P_M-qA_M)(P_N-qA_N)+m^2e^{-2\phi}=0.
\label{massshell}
\end{equation}

To obtain the expression for the Hamiltonian $H=-P_t$ for the probe 
D0-brane mechanics, we solve this mass-shell constraint (\ref{massshell}).  
We consider the following general spherically symmetric $D=10$ metric 
Ansatz:
\begin{eqnarray}
G_{MN}dx^Mdx^N&=&-A(r)dt^2+B(r)dr^2+C(r)d\Omega^2_8
\cr
&=&-A(r)dt^2+B(r)dr^2+C(r)[d\theta^2+\cos^2\theta d\psi^2_1+
\cos^2\theta\cos^2\psi_1d\psi^2_2
\cr
& &+\cos^2\theta\cos^2\psi_1\cos^2\psi_2d\psi^2_3+\mu^2_id\phi^2_i],
\label{sphansatz}
\end{eqnarray}
where 
\begin{equation}
\mu_1=\sin\theta,\ \   
\mu_2=\cos\theta\sin\psi_1,\ \ 
\mu_3=\cos\theta\cos\psi_1\sin\psi_2,\ \ 
\mu_4=\cos\theta\cos\psi_1\cos\psi_2\sin\psi_3.
\label{mudefs}
\end{equation}
Then, the expression for the Hamiltonian $H$ takes the following form:
\begin{equation}
H={{P^2_r}\over{2f}}+{{g}\over{2f}},
\label{genhamilton}
\end{equation}
where
\begin{eqnarray}
f&\equiv&{1\over 2}A^{-{1\over 2}}Be^{-\phi}\left[\sqrt{m^2+e^{2\phi}
(P^2_r+BC^{-1}\vec{L}^2)/B}+qA^{-{1\over 2}}A_te^{\phi}\right],
\cr
g&\equiv&Be^{-2\phi}\left[(m^2-q^2A^{-1}A^2_te^{2\phi})+C^{-1}
e^{2\phi}\vec{L}^2\right].
\label{deffg}
\end{eqnarray}
Here, $A_t$ is the time component of the 1-form field $A_M$ of the 
near-horizon source D0-brane supergravity solution (\ref{sgdecoupl}) 
and $\vec{L}^2$ is the angular momentum operator of the probe D0-brane 
given by
\begin{equation}
\vec{L}^2=P^2_{\theta}+{{P^2_{\psi_1}}\over{\cos^2\theta}}+
{{P^2_{\psi_2}}\over{\cos^2\theta\cos^2\psi_1}}+
{{P^2_{\psi_3}}\over{\cos^2\theta\cos^2\psi_1\cos^2\psi_2}}+
\sum^4_{i=1}{{P^2_{\phi_i}}\over{\mu^2_i}}.
\label{angop}
\end{equation}

First, we consider the probe D0-brane moving in the string-frame near 
horizon background of the source D0-brane.  In the original work 
\cite{kall1} of the superconformal mechanics of a test charged 
particle in the near horizon geometry of the Reissner-Nordstr\"om 
black hole, it was necessary to redefine the radial coordinate so 
that the metric components $A$ and $B$ in Eq. (\ref{sphansatz}) 
satisfy the relation $A=B^2$ for the purpose of setting the factor 
$A^{-{1\over 2}}B$ in Eq. (\ref{genhamilton}) equal to 1.  However, 
in the D0-brane case, as we will see, it is more convenient to work 
with the original form (\ref{sgdecoupl}) of near horizon metric, 
since the expression for the Hamiltonian becomes simpler in the 
original radial coordinate.  

By substituting the string-frame near horizon metric into the 
general formulae (\ref{genhamilton}) and (\ref{deffg}), one obtains 
the following Hamiltonian for the probe D0-brane in the string-frame 
near horizon background of the source D0-brane bound state:
\begin{equation}
H={{P^2_r}\over{2f}}+{{g}\over{2f}},
\label{stringharm}
\end{equation}
where $f$ and $g$ are given by 
\begin{eqnarray}
f&=&{1\over 2}g^{-1}_s\left[\sqrt{m^2+g^2_s\left({\mu\over r}\right)^7
\left(P^2_r+{{\vec{L}^2}\over{r^2}}\right)}+q\right],
\cr
g&=&g^{-2}_s\left({\mu\over r}\right)^{-7}
(m^2-q^2)+{{\vec{L}^2}\over {r^2}},
\label{gfforstring}
\end{eqnarray}
or in terms of the SYM theory variables
\begin{equation}
H={{P^2_U}\over{2f}}+{{g}\over{2f}},
\label{stringharmym}
\end{equation}
where $f$ and $g$ are given by 
\begin{eqnarray}
f&=&{1\over 2}\alpha^{\prime\,{1\over 2}}g^{-2}_{YM}
\left[\sqrt{m^2+{{\alpha^{\prime\,-1}g^6_{YM}N}\over{U^7}}
\left(P^2_U+{{\vec{L}^2}\over{U^2}}\right)}+q\right],
\cr
g&=&\left({{\alpha^{\prime\,-1}g^6_{YM}N}\over{U^7}}\right)^{-1}
(m^2-q^2)+{{\vec{L}^2}\over{U^2}}.
\label{gfforstringym}
\end{eqnarray}
It is interesting that the term $A^{-1}A^2_te^{2\phi}$ in Eq. (\ref{deffg}) 
becomes 1 for the near horizon D0-brane solution (\ref{sgdecoupl}). 
So, the expressions for $f$ and $g$ becomes greatly simplified.  
And, in particular, in the extreme limit ($m-q\to 0$) of the probe D0-brane, 
the first term in $g$ drops out.  This also generally holds for any dilatonic 
0-brane supergravity solutions.

Just as in the case of the test charged particle in the Reissner-Nordsr\"om 
black hole background, the mechanics of the probe D0-brane has the 
$SL(2,{\bf R})$ symmetry with the following generators:
\begin{equation}
H={{P^2_r}\over{2f}}+{g\over{2f}},\ \ \ \ 
K=-{1\over 2}fr^2,\ \ \ \ 
D={1\over 2}rP_r,
\label{sl2rgen}
\end{equation}
where the Hamiltonian $H$ generates the time translation, 
$K$ generates the special conformal transformation and 
$D$ generates the scale transformation or the dilatation. 
These generators satisfy the following $SL(2,{\bf R})$ algebra:
\begin{equation}
[D,H]=H,\ \ \ \ [D,H]=-K,\ \ \ \ [H,K]=2D.
\label{sl2ralg}
\end{equation}
This is the D0-brane generalization of conformal quantum mechanics studied 
in Refs. \cite{scm1,scm2,kall1}.  

In fact, the near horizon solution (\ref{sgdecoupl}), when uplifted 
as a solution of the 11-dimensional gravity (i.e. the 11-dimensional 
plane wave solution), is invariant under the $SU(1,1)\cong SL(2,{\bf R})$ 
isometry
\footnote{At the 10-dimensional level, such $SL(2,{\bf R})$ 
transformations act on the near-horizon supergravity D0-brane solution 
in such a way that the constant $Q$ in the harmonic function $H$ 
transforms as if it is a ``field'' on the worldvolume and then is set 
to a constant after the transformations \cite{jevi2}.} 
generated by the scale transformation $\delta_D$, the special 
coordinate transformation $\delta_K$ and the time translation 
$\delta_H$ \cite{jevi1}.  Under the time translation, $\delta_Ht=1$, 
$\delta_HU=0$ and $\delta_Hg_s=0$.  Under the special coordinate 
transformation, $\delta_Kt=-(t^2+k{{g^2_{YM}}\over{U^5}})$, 
$\delta_KU=2tU$ and  $\delta_Kg_s=6tg_s$.  Finally, under the dilatation, 
$\delta_Dt=-t$, $\delta_DU=U$ and $\delta_Dg_s=3g_s$.  And $A_t$ transforms 
as a conformal field of dimension 1.  These infinitesimal transformations 
satisfy the following $SL(2,{\bf R})$ algebra just like the symmetry 
generators (\ref{sl2rgen}) of the probe D0-brane mechanics:
\begin{equation}
[\delta_D,\delta_H]=\delta_H,\ \ \ \ 
[\delta_D,\delta_K]=-\delta_K,\ \ \ \ 
[\delta_H,\delta_K]=2\delta_D.
\label{sl2r11d}
\end{equation}

Note, the string coupling $g_s$ changes under the dilatation and the 
special coordinate transformation, and especially $g_s$ becomes time 
dependent after the special coordinate transformation is applied.  
Thereby, this $SL(2,{\bf R})$ isometry of the near horizon geometry 
does not extend to a conformal symmetry of the complete supergravity 
solution.  The corresponding boundary theory, i.e. $0+1$ dimensional 
SYM matrix quantum mechanics, has the same $SL(2,{\bf R})$ symmetry, 
but the string coupling $g_s$ also transforms under this symmetry, 
unlike the case of D3-brane.  However, as noted in Ref. \cite{jevi1}, 
since the dilaton coupling $g_s$ is related to the matrix model coupling 
constant, one can still think of `generalized' $SL(2,{\bf R})$ conformal 
symmetry in which the string coupling is now regarded as a part of 
background fields that transform under the symmetry.  The `generalized' 
conformal symmetry therefore transforms a matrix model at one value of 
the coupling constant to another.  Note, as pointed out in the previous 
section, D0-brane in the dual frame has description in terms of domain-wall 
solution after the compactification on $S^8$.  So, this is also related to 
the fact that in the domain-wall/QFT correspondence the choice of horosphere 
(the hypersurface of constant $u$) for the Minkowski vacuum corresponds 
to a choice of coupling constant of a non-conformal QFT \cite{town3}.  
In the non-conformal case, the interpolation between the AdS Killing horizon 
(in the dual frame) and its boundary therefore corresponds to an 
interpolation between strong and weak coupling, or vice versa.  

The extreme limit of the probe D0-brane ($(m-q)\to 0$) can be interpreted 
as the M theory in the LCF, since in this case both the source and the probe 
D0-branes are in the BPS limit.  In this case, the functions $f$ and $g$ are 
simplified to
\begin{equation}
f={1\over 2}g^{-1}_s\left[\sqrt{m^2+g^2_s\left({\mu\over r}\right)^7
\left(P^2_r+{{\vec{L}^2}\over{r^2}}\right)}+m\right],\ \ \ \ 
g={{\vec{L}^2}\over {r^2}},
\label{extfgstr}
\end{equation}
or in terms of the SYM theory variables
\begin{equation}
f={1\over 2}\alpha^{\prime\,{1\over 2}}g^{-2}_{YM}
\left[\sqrt{m^2+{{\alpha^{\prime\,-1}g^6_{YM}N}\over{U^7}}
\left(P^2_U+{{\vec{L}^2}\over{U^2}}\right)}+m\right],\ \ \ \ 
g={{\vec{L}^2}\over{U^2}}.
\label{extfgstrym}
\end{equation}

Unlike the case of a charged test particle in the near horizon 
Reissner-Nordsr\"om black hole background studied in Ref. \cite{kall1}, 
we do not let the mass of the source D0-brane bound state go to 
infinity, since the number $N$ of the D0-branes in the M theory 
on the LCF is kept finite.  The fact that we are considering the 
near horizon geometry of the D0-brane supergravity solution means 
that we are in the LCF of M theory, since the near horizon geometry 
of the D0-brane supergravity solution is also the null reduction of 
the M wave supergravity solution \cite{tsey1}, which is interpreted as 
M theory in the LCF. Taking $N$ to infinity, i.e. infinitely massive 
source D0-brane, corresponds to M theory in the IMF.  Note, the IMF 
is defined as the limit in which $N$ and $R_{11}$ go to infinity 
such that the momentum $P_{11}=N/R_{11}$ also goes to infinity.  Since 
$P_{11}\sim g^{-7/2}_{YM}N^{1/4}U^{21/4}$ and $R_{11}\sim g^{7/2}_{YM}
N^{3/4}U^{-21/4}$, one can take both $P_{11}$ and $R_{11}$ to 
infinity while taking $N\to\infty$, if $U$ is in the range of 
$g^{2/3}_{YM}N^{-1/21}\ll U\ll g^{2/3}_{YM}N^{1/7}$.  However, this 
range of $U$ is beyond the range of validity (\ref{rcnstrnt}) of the 
near horizon solution (\ref{sgdecoupl}).  In other words, one cannot 
go to the IMF of M theory while keeping the parameters within the 
validity of the near horizon solution (\ref{sgdecoupl}).  Anyway, the 
IMF means decompactification ($R_{11}\to\infty$) to the eleven 
dimensions.  So, we should rather consider the M wave solution in the 
case $N\to\infty$.  

By expanding this Hamiltonian for an extreme $D0$-brane probe, one 
obtains a Hamiltonian of the form which is the sum of the non-relativistic 
kinetic term for the probe and the velocity dependent potential given by 
the sum of terms of the form $\sim {{v^{2n+2}}\over{r^{7n}}}$ 
($n\in{\bf Z}^+$).  This should reproduce the (M)atrix theory calculation 
of graviton-graviton scattering and its supergravity calculation 
reproduction of Ref. \cite{tsey1}.  This is because the above action 
(\ref{dpartact}) is just the reduction of the action for the probe 
in the background of $D=11$ plane wave (describing the motion of the 
probe graviton in the background of the moving heavy source graviton) on 
a light-like circle to 10 dimensions, i.e. M theory in the LCF.  
Note also that since M theory in the LCF has ``Galilean'' invariance (or is 
described by non-relativistic quantum mechanics) in the transverse space, 
the above Hamiltonian for an extreme probe D0-brane has the non-relativistic 
structure.  These above arguments, together with the fact that the 
generalized conformal mechanics of probe D0-brane and the boundary $SU(N)$ 
SYM theory (i.e. the (M)atrix model of M theory in the LCF) satisfy the 
same $SL(2,{\bf R})$ symmetry, implies the equivalence between the 
(M)atrix model with finite $N$ and the generalized conformal mechanics 
of D0-brane.  

Next, we consider the probe D0-brane moving in the background of the 
dual-frame metric of the source D0-brane bound state.  Since the dual 
frame can be considered as a ``holographic frame'' describing supergravity 
probes \cite{town3}, it is worthwhile to consider the case of the dual frame.  
In this frame, the near horizon metric takes the AdS$_2\times S^8$ form.  
So, one should expect that the `generalized' conformal mechanics with the 
$SL(2,{\bf R})$ symmetry can also be realized in the dual frame.  The 
action for the probe D0-brane in the dual frame background of the source 
D0-brane is given by (\ref{dpartact}) with the string-frame metric 
$G^{str}_{MN}$ replaced by the dual-frame metric $G^{dual}_{MN}$ through 
the relation $G^{str}_{MN}=e^{{2\over 7}\phi}G^{dual}_{MN}$.  So, for the 
probe D0-brane in the dual frame, the dilaton factor $e^{-\phi}$ in Eqs. 
(\ref{dpartact}) and (\ref{conjmom}) [the dilaton factor $e^{-2\phi}$ in 
Eqs. (\ref{massshell}) and (\ref{deffg})] is replaced by $e^{-{6\over 7}
\phi}$ [$e^{-{{12}\over 7}\phi}$].   

Substituting the dual-frame near horizon solution (\ref{dualmet}) into 
this general expression for the Hamiltonian corresponding to the dual 
frame, one finds that the Hamiltonian in the dual frame has the same form 
as the string-frame Hamiltonian (\ref{stringharm}) with the same $f$ and 
$g$ (\ref{gfforstring}).  In fact, in general the Hamiltonian 
describing probe D0-brane is independent of the near horizon spacetime 
frame of the source D0-brane.  So, the dynamics of the probe D0-brane in 
the dual frame also has the $SL(2,{\bf R})$ symmetry with the same 
symmetry generators (\ref{sl2rgen}) and algebra (\ref{sl2ralg}) as the 
string-frame case.  

As pointed out in the previous section, the source D0-brane solution 
also has the (Hodge) dual description in terms of the 8-form field 
strength, whose magnetic charge now is carried by the source D0-brane.  
In this case, by compactifying the D0-brane solution on $S^8$ one obtains 
a domain wall solution in $1+1$ dimensions \cite{town4,town3}.  This domain 
wall solution solves the equations of motion of a $D=2$ $SO(9)$ gauged 
maximal supergravity theory, which is an $S^8$ compactification of the 
type IIA supergravity.  This $SO(9)$, which is the largest subgroup 
of the $SO(16)$ and the isometry group of $S^8$ upon which the D0-brane is 
compactified, is also the R-symmetry group of the corresponding boundary 
$D=1$ QFT.  In general, the R-symmetry of the supersymmetric QFT on the 
domain wall worldvolume matches the gauge group, which is the isometry 
group of the compactification manifold, of the equivalent gauged 
supergravity \cite{town3}.  

Putting together all the above facts, namely ($i$) generalized 
conformal mechanics of the probe D0-brane in the string-frame 
near horizon background of the source D0-brane is related to 
the M theory in the LCF, ($ii$) the generalized conformal mechanics 
of the probe D0-branes in the string and the dual frames of the 
near horizon source D0-branes are described by the same Hamiltonian, 
($iii$) upon the dimensional reduction on $S^8$ the bulk theory of 
the source D0-brane in the dual frame in the Hodge-dual description 
(in terms of the 8-form field strength) is the $D=2$ $SO(9)$ gauged 
maximal supergravity theory, one arrives at the speculation that 
the M theory in the LCF is related to the $D=2$ $SO(9)$ gauged 
maximal supergravity theory (i.e. a $D=2$ Kaluza-Klein supergravity 
theory with domain wall vacuum).

\end{document}